\documentclass[a4paper,fleqn,usenatbib]{mnras}

\usepackage{mathtools}  
\usepackage{amssymb}
\usepackage[T1]{fontenc}
\usepackage{ae,aecompl}
\usepackage{adjustbox}

\usepackage{graphicx}	
\usepackage{amsmath}	
\usepackage{amssymb}	
\usepackage{natbib} 

\title[Polarimetry with the K~{\sc i} lines]{Solar polarimetry through the K~{\sc i} lines at 770 nm}

\author[C. Quintero Noda et al.]{C. Quintero Noda,$^{1}$\thanks{E-mail: carlos@solar.isas.jaxa.jp}
H. Uitenbroek,$^{2}$
Y. Katsukawa,$^{3}$
T. Shimizu,$^{1}$
T. Oba,$^{4}$
\newauthor
M. Carlsson,$^{5}$
D. Orozco Su\'arez,$^{6}$
B. Ruiz Cobo,$^{7,8}$
M. Kubo,$^{3}$
T. Anan,$^{9}$
\newauthor
K. Ichimoto,$^{3,9}$
Y. Suematsu$^{3}$ 
\\
$^{1}$Institute of Space and Astronautical Science, Japan Aerospace Exploration Agency, Sagamihara, Kanagawa 252-5210, Japan\\
$^{2}$National Solar Observatory, University of Colorado Boulder, 3665 Discovery Drive, Boulder, CO 80303, USA\\
$^{3}$National Astronomical Observatory of Japan, 2-21-1 Osawa, Mitaka, Tokyo 181-8588, Japan\\
$^{4}$SOKENDAI, Shonan Village, Hayama, Kanagawa 240-0193 Japan\\
$^{5}$Institute of Theoretical Astrophysics, University of Oslo, P.O. Box 1029 Blindern, N-0315 Oslo, Norway\\
$^{6}$Instituto de Astrof\'isica de Andaluc\'ia (CSIC), Glorieta de la Astronom\'ia, 18008 Granada, Spain\\
$^{7}$Instituto de Astrof\'isica de Canarias, E-38200, La Laguna, Tenerife, Spain.\\
$^{8}$Departamento de Astrof\'isica, Univ. de La Laguna, La Laguna, Tenerife, E-38205, Spain\\
$^{9}$Kwasan and Hida Observatories, Kyoto University, Kurabashira Kamitakara-cho, Takayama-city, 506-1314 Gifu, Japan\\
}

\date{Accepted XXX. Received YYY; in original form ZZZ}

\pubyear{2017}

\begin{document}
\label{firstpage}
\pagerange{\pageref{firstpage}--\pageref{lastpage}}
\maketitle

\begin{abstract}
We characterize the K~{\sc i} $\rm D_1$ \& $\rm D_2$ lines in order to determine  whether they could complement the 850~nm window, containing the Ca~{\sc ii} infrared triplet lines and several Zeeman sensitive photospheric lines, that was studied previously. We investigate the effect of partial redistribution on the intensity profiles, their sensitivity to changes in different atmospheric parameters, and the spatial distribution of Zeeman polarization signals employing a realistic magnetohydrodynamic simulation. The results show that these lines form in the upper photosphere at around 500~km and that they are sensitive to the line of sight velocity and magnetic field strength at heights where neither the photospheric lines nor the Ca~{\sc ii} infrared lines are. However, at the same time, we found that their sensitivity to the temperature essentially comes from the photosphere. Then, we conclude that the K~{\sc i} lines provide a complement to the lines in the 850~nm window for the determination of atmospheric parameters in the upper photosphere, especially for the line of sight velocity and the magnetic field.
\end{abstract}

\begin{keywords}
Sun: chromosphere -- Sun: magnetic fields -- techniques: polarimetric
\end{keywords}



\section{Introduction}

Future solar missions aim to understand the solar atmosphere through the characterization of  the magnetic field and its role in different solar phenomena. At present, we have access to routine polarimetric observations, e.g., Hinode/SP \citep{Tsuneta2008,Lites2013} or SDO/HMI  \citep{Pesnell2012,Schou2012}, that allow to infer the properties of the magnetic field but only from the lower part of the solar atmosphere, i.e. the photosphere. Meanwhile, polarimetric observations of the chromosphere are scarce (see the works using observations performed with  SST/CRISP \citep{Scharmer2003,Scharmer2008} or DST/IBIS \citep{Cavallini2006} among others), while for the case of the corona they are extremely limited \citep[for example,][]{Lin2004,Tomczyk2008}. In this regard, future missions as Solar-C \citep[][]{Katsukawa2011,Katsukawa2012,Watanabe2014,Suematsu2016} and ground-based telescopes as DKIST \citep{Elmore2014} or EST \citep{Collados2013} aim to infer the magnetic field in the chromosphere, the most accessible one from the two mentioned layers (although the former aims to observe the corona too). The target is to understand some of the physical phenomena related to, for example, the transfer of energy from lower to upper layers in the form of jets \citep{Shibata2007} or travelling waves \citep{dePontieu2007}. In both cases, the magnetic field plays a crucial role as trigger mechanism or guiding channel although we barely have any information from it above the photosphere. Therefore, we have to perform polarimetric observations of spectral lines that form in the chromosphere if we want to be closer to solve some of the long-standing questions of solar physics. 

We started in \cite{QuinteroNoda2016} studying the chromospheric Ca~{\sc ii} 8542 \AA \ line that has been extensively observed with imaging instruments such as SST/CRISP and DST/IBIS but also with some spectrographs as DST/SPINOR \citep{SocasNavarro2006}, among others. We concluded in that work that the line gives information about the lower to mid chromosphere and is very sensitive to the temperature and line of sight (LOS) velocity at all layers from the photosphere to the chromosphere. Unfortunately, the sensitivity of the calcium triplet lines to the magnetic field vector through the Zeeman effect, in particular to its transversal component,  is weak at lower heights. This fact motivated our second work related to the characterization of chromospheric lines, i.e. \cite{QuinteroNoda2017}, where we looked for spectral lines in the vicinity of the Ca~{\sc ii} 8542~\AA \ line trying to find a candidate with higher sensitivity to the magnetic field in the photosphere. Interestingly, we found several photospheric lines that are very promising and are also located closer to a second spectral line from the Ca~{\sc ii} infrared triplet, i.e. the Ca~{\sc ii} 8498 \AA \ line. These findings demonstrate that the spectral region located at 850~nm is one of the best candidates for performing simultaneous polarimetric observations of the photosphere and the chromosphere. However, in the present work we want to address a different topic that is related to the low sensitivity at a height around the temperature minimum sampled by the 850~nm spectral window \citep[see the deep valley in the response functions (RF) of Figure 4 of][]{QuinteroNoda2017}. In this regard, we would like to examine spectral lines that form in the upper photosphere, i.e. between approximately $450-600$~km above the height where the continuum optical depth $\tau_{500}$ is unity, as the K~{\sc i} $\rm D_1$ \& $\rm D_2$ lines. These lines have been scarcely explored as they are located in a spectral region strongly affected by the atmospheric O$_{2}$ $A$-band \citep[for example, ][]{Babcock1948}. In fact, the K~{\sc i} $\rm D_2$ line is completely blocked by one of these O$_{2}$ lines \citep[see, for instance, the solar atlas of][]{Delbouille1973} making it impossible to observe it from the ground. Therefore, these lines are appealing for future space missions and balloon missions as the {\sc sunrise} balloon project \citep{Solanki2010,Barthol2011,Solanki2017}.

\section{Spectral lines}

The K~{\sc i} $\rm D_1$ \& $\rm D_2$ are two resonance lines of neutral potassium located at 7698.97 and 7664.90~\AA, respectively. These lines have been theoretically studied in the past, for instance, in the series ``The formation of helioseismology lines'' of \cite{Bruls1992a,Bruls1992b,Uitenbroek1992}, due to the popularity of alkali lines for helioseismology studies using resonance cells \citep{Brookes1978} in the decade of 1980s \citep[e.g.,~][]{Claverie1982,Jefferies1988,Palle1989}. From the mentioned series of theoretical works, we know that the K~{\sc i} $\rm D_1$ (the $\rm D_2$ line is not studied in there as it is blended with an O$_{2}$ line) forms around $500\sim600$~km above the solar surface and non$-$local thermodynamic equilibrium is required to compute the atomic populations of the levels involved in the transition, although partial redistribution effects (PRD) are not important \citep{Uitenbroek1992}.

Unfortunately, after those studies, the popularity of the potassium lines for the estimation of solar atmospheric parameters, in particular magnetic fields, did not increase, although some works as the one from \cite{Jefferies2006} can be found in the literature. In fact, from the alkali lines, the Na~{\sc i} is presently observed \citep[some recent examples can be found in][]{Pietarila2010,Rutten2011,Rutten2015} and also studied with 3D realistic simulations \citep{Leenaarts2010}. Moreover, there are also works that study its ``enigmatic'' atomic polarization \citep[e.g., ][]{Belluzzi2015} that can be found in quiet Sun observations \citep[e.g.][]{Stenflo1997} of the ``second solar spectrum'' \citep{Ivanov1991}. We believe that the reason behind the lack of studies regarding the K~{\sc i} $\rm D_1$ \& $\rm D_2$ lines is that the spectrum is highly dominated by O$_{2}$ lines that preclude the interpretation of observations of those lines. Something that is not related to the intrinsic capabilities of the K~{\sc i} lines for inferring atmospheric information from the upper photosphere.

Besides the two K~{\sc i} lines,  we also inspect neighbouring spectral lines trying to find promising photospheric candidates. We restrict the wavelength coverage to a value of approximately $60$~\AA \ in order to be able to fit it in a hypothetical camera of 2K$\times$2K pixels while maintaining a moderate/high spectral resolution. We include in Table~\ref{lines} all the spectral lines we consider as good candidates. Their atomic information was retrieved from the database of R.~Kurucz \citep{Kurucz1995}. In addition, for the case of the K~{\sc i} lines, we include the oscillator strength from the atomic model used in this work. Those values are similar to the ones given in \cite{Wiese1969} (see page 228 of the monograph). The mentioned atomic model contains 12 levels plus the continuum and was presented in \cite{Bruls1992a} as a simplified version of the comprehensive model. However, this one is complex enough to produce similar intensity profiles in most of the cases (see Figure 21 of the mentioned work). Additionally, we use the original photo-ionizations and collisions values included in the mentioned atomic model. Finally, the abundance of K~{\sc i} we utilize is equal to 5.12 and was extracted from \cite{Grevesse1998}.

\begin{table}
\hspace{-0.5cm}
\normalsize
\begin{adjustbox}{width=0.495\textwidth}
  \bgroup
\def\arraystretch{1.25}
\begin{tabular}{lccccccc}
	\hline
 & 	Atom                   & $\lambda$ [\AA] & $\log gf$ & $L_l$        & $U_l$            & $\rm g_{eff}$ & $\rm I_{core}$ [au] \\
	\hline
1      & Mg~{\sc i}             &  7657.60      & -1.120  & ${}^3S_{1}$      &  ${}^3P^{0}_{2}$      & 1.25  & 4917    \\
2      & Mg~{\sc i}             &  7659.15      & -1.340  & ${}^3S_{1}$      &  ${}^3P^{0}_{1}$      & 1.75  & 6736    \\
3      & Fe~{\sc i}             &  7661.20      & -0.914  & ${}^5F^{0}_{3}$  &  ${}^5F_{4}$          & 1.50   & 4791    \\
4      & Fe~{\sc i}             &  7664.30      & -1.682  & ${}^3G_{4}$      &  ${}^3F^{0}_{3}$      & 1.00   & 3941    \\
5      & K~{\sc i}              &  7664.90      & 0.1345  & ${}^2S_{1/2}$    &  ${}^2P^{0}_{3/2}$    & 1.16  & -----    \\
6      & Si~{\sc i}             &  7680.25      & -0.690  & ${}^1P_{1}$      &  ${}^1D^{0}_{2}$      & 1.00   & 5488    \\
7      & Mg~{\sc i}             &  7691.55      & -0.800  & ${}^1D_{2}$      &  ${}^1F^{0}_{3}$      & 1.00   & 5568    \\
8      & K~{\sc i}              &  7698.97      & -0.169  & ${}^2S_{1/2}$    &  ${}^2P^{0}_{1/2}$    & 1.33  & 1663    \\
9      & Fe~{\sc i}             &  7710.36      & -1.113  & ${}^5F^{0}_{4}$  &  ${}^5F_{5}$          & 1.50   & 5152    \\
10     & Ni~{\sc i}             &  7714.32      & -2.200  & ${}^3P_{2}$      &  ${}^3P^{0}_{2}$      & 1.50   & 4132    \\
	\hline
  \end{tabular}
  \egroup
\end{adjustbox}
\caption {Spectral lines included in the infrared 770~nm window. Each column, from left to right, contains the number assigned to each line (see also Figure \ref{atlas}), the corresponding atomic species, line core wavelength, $\log gf$ of the transition, the spectroscopic notation of the lower and the upper level, the effective Land\'{e} factor, and the line core intensity in arbitrary units (the continuum level corresponds to 10000).} \label{lines}     
\end{table}

\begin{figure}
\begin{center} 
 \includegraphics[trim=35 0 +25 0,width=8.0cm]{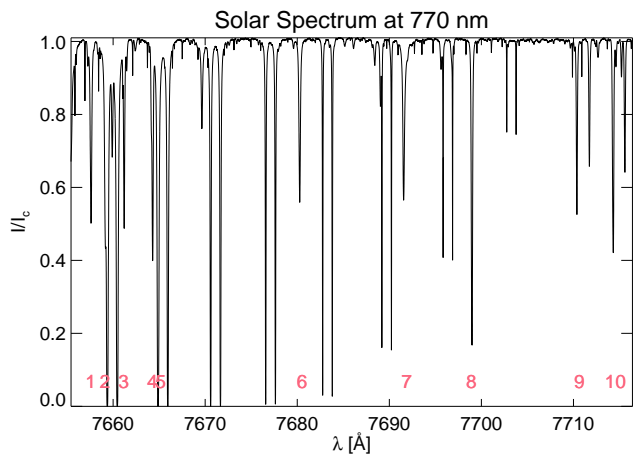}
 \vspace{-0.3cm}
 \caption{The observed solar atlas of the spectral region around 770~nm. We indicate with numbers some of the lines that are included in this spectrum and described in Table \ref{lines}. The K~{\sc i}~$\rm D_1$ \& $\rm D_2$ lines are of particular interest with labels 8 and 5, respectively.}
 \label{atlas}
 \end{center}
\end{figure}

\begin{figure*}
\begin{center} 
 \includegraphics[trim=-6 0 20 0,width=17.0cm]{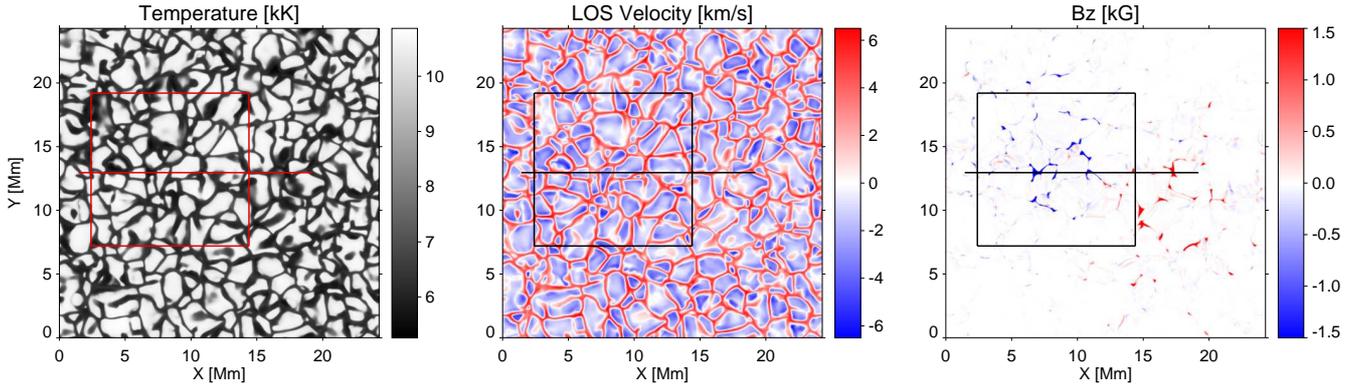}
 \vspace{-0.1cm}
 \caption{Snapshot 385 from the {\sc bifrost} enhanced network simulation. From left to right, temperature, LOS velocity, and longitudinal field strength at the geometrical height $Z=0$~km. Two regions used in this work are shown with a horizontal line and a squared box.}
 \label{fov}
 \end{center}
\end{figure*}

We plot in Figure~\ref{atlas} the solar atlas  observed by \cite{Delbouille1973} where we indicate with numbers the location of the lines presented in Table~\ref{lines}. We can see that there are several candidates with moderate effective Land\'{e} factors (see numbers 2, 3, 9, and 10) although none of them are as strong as the Fe~{\sc i} 8468 and 8514 \AA \ lines \citep[see Table 1 of ][]{QuinteroNoda2017}. In addition, we have the K~{\sc i} $\rm D_1$ \& $\rm D_2$ lines designated by labels 8 and 5, respectively. In the case of the latter, there is a photospheric line close to it, i.e. label 4. Moreover, we can see that the spectrum is dominated by O$_{2}$ lines that absorb (sometimes completely) the solar radiation, what indeed impedes the observation of the K~{\sc i} $\rm D_2$ line from the ground.

\begin{figure*}
\begin{center} 
 \includegraphics[trim=20 0 20 0,width=15.0cm]{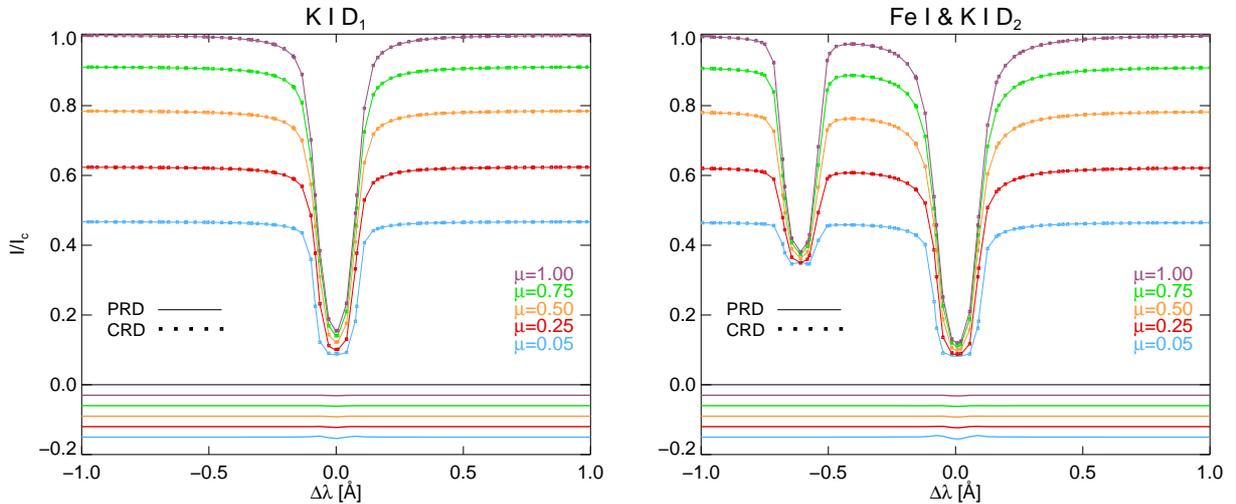}
 \vspace{+0.2cm}
 \caption{PRD (solid) and CRD (squares) profiles for the K~{\sc i} $\rm D_1$ \& $\rm D_2$ lines (we also included the neighbour Fe~{\sc i} line for the latter case). Colours indicate the selected heliocentric angle (larger $\mu$ corresponds to a line of sight closer to the disc centre and to higher continuum intensity). In addition, we show the difference between PRD and CRD profiles in the bottom of each panel.}
 \label{prd1d}
 \end{center}
\end{figure*}

\section{Methodology}

In the present work we employ the RH code \citep{Uitenbroek2001,Uitenbroek2003} to synthesize the Stokes profiles. One of our targets is to check the effect of PRD on the K~{\sc i}~$\rm D_2$ line (for the first time) and to revisit the results for the K~{\sc i}~$\rm D_1$ line using additional atmospheric models. The code essentially works in two modes, computing the intensity spectrum with a variety of options (including PRD), or computing the full stokes vector where only complete redistribution (CRD) is available. We start this work computing the intensity profiles for different scenarios under both regimes, i.e. PRD and CRD. This is because, although \cite{Uitenbroek1992} established that the K~{\sc i}~$\rm D_1$ line was not sensitive to PRD effects, nothing was said about the K~{\sc i}~$\rm D_2$ line. After that, we compute the Stokes vector under the CRD approximation.

We use two types of atmospheric models. First, we start with the semi-empirical FALC model \citep{Fontenla1993}. Later, we use the snapshot 385 of the {\sc bifrost} \citep{Gudiksen2011}  enhanced network simulation \citep{Carlsson2016} that has been used in several works related to the synthesis of chromospheric spectra \citep[e.g.,][]{delaCruzRodriguez2012,delaCruzRodriguez2013,Leenaarts2015,Stepan2016,Sukhorukov2017}, including our studies of the 850~nm spectral window. We show the spatial distribution of several atmospheric parameters at a geometrical height $Z=0$~km for the computed full field of view in Figure~\ref{fov}. We highlight two different regions. First, the horizontal line designates the region we used for the 2D studies we want to perform and second, the squared region delimits the field of view that will be used for additional studies in the latter part of this work. Regarding the properties of this simulation, we refer the reader to the work of \cite{Carlsson2016}. We do not include any spatial degradation in our studies, i.e. we use the original horizontal pixel size of 48~km. We add an artificial microturbulence of 3~km/s constant with height as we did in previous works. However, in the future we should use a microturbulence that changes with height as 3~km/s is higher than what it is expected in the photosphere \citep[see, for instance, the FALC model of ][]{Fontenla1993}.

Finally, as one of the targets of these studies is to examine if the K~{\sc i} lines can help us to surpass the limitations of the 850~nm  window, we compare the results with those from selected lines of that region. In that sense, we repeat all the computations for the latter window using RH assuming CRD and disc centre observations, i.e.  $\mu=1$, where $\mu=\cos(\theta)$ and $\theta$ the heliocentric angle. We use the same atomic parameters presented in Table 1 of \cite{QuinteroNoda2017} and the same atomic model of 5 levels plus continuum for the two infrared Ca~{\sc ii} lines \citep[see][]{Shine1974}.

\begin{figure*}
\begin{center} 
 \includegraphics[trim=15 0 28 0,width=15.4cm]{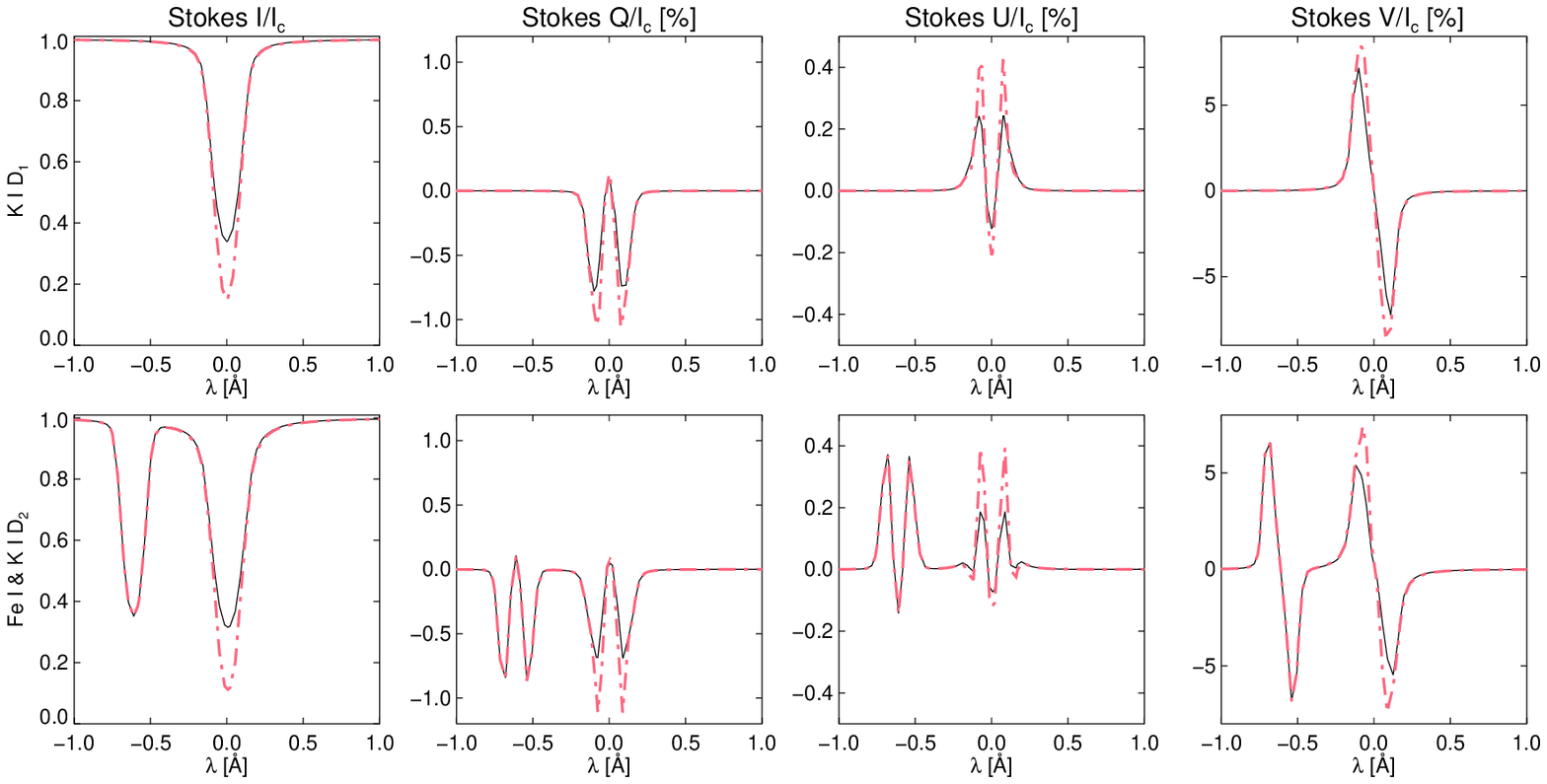}
 \vspace{+0.1cm}
 \caption{Comparison between a NLTE (dashed line) and an LTE (solid line) computation for the K~{\sc i} $\rm D_1$ \& $\rm D_2$ lines (we also included the neighbour Fe~{\sc i} line, always computed in LTE, for the latter case). We use the FALC atmosphere adding a magnetic field of 500~G, 45~degrees of inclination and 70~degrees of azimuth, constant with height.}
 \label{nlte}
 \end{center}
\end{figure*}

\begin{figure}
\begin{center} 
 \includegraphics[trim=0 0 0 0,width=6.6cm]{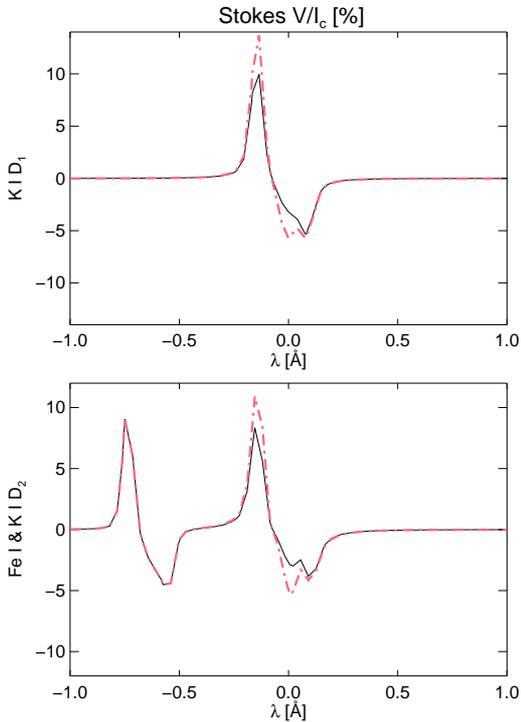}
 \vspace{+0.1cm}
 \caption{Same as Figure~\ref{nlte} but we only focus on the Stokes $V$ profiles this time. We include a gradient in the velocity stratification in order to produce an area and amplitude asymmetry in the profiles.}
 \label{nlte_asymm}
 \end{center}
\end{figure}

\begin{figure*}
\begin{center} 
 \includegraphics[trim=25 0 10 0,width=15.3cm]{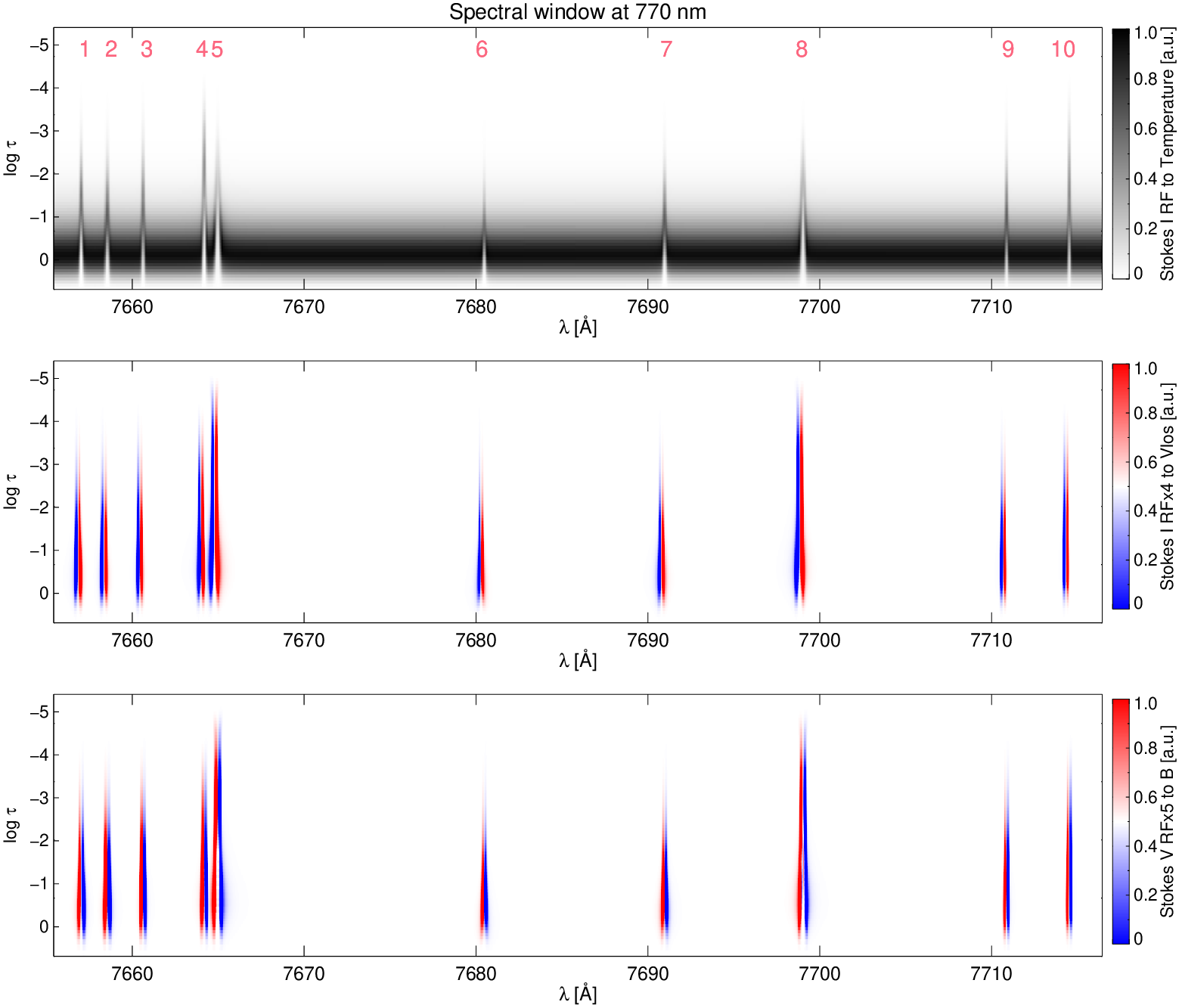}
 \vspace{-0.0cm}
 \caption{Response functions to changes in the temperature (top), LOS velocity (middle), and field strength (bottom). White colour indicates no sensitivity to changes in the atmospheric parameters, while colour (or black) means that the spectral lines are sensitive to a perturbation of a given atmospheric parameter at a given height. We use the FALC atmospheric model and all the cases are normalized to the maximum of the Stokes $I$ RF to changes in the temperature (top panel).}
 \label{2drf}
 \end{center}
\end{figure*}

\subsection{Partial redistribution effects}

During a scattering process, photons are absorbed by an atom, modifying the population of the atomic levels, and then they are re-emitted. If the frequency of both photons is fully correlated, we are under coherent scattering, while if there is no relation at all, we are under complete redistribution \cite[see, for example, the monograph of ][]{Jefferies1968}. However, in general, photon redistribution during scattering represents a combination of the mentioned regimes called partial frequency redistribution. In the case of solar resonance lines (the lifetime of the lower level is very large), it is well known that they are affected by coherent scattering effects \citep[some works on this topic are][]{Uitenbroek1989,Uitenbroek1992,Belluzzi2014,delPinoAleman2016}. However, in the case of the K~{\sc i} lines it seems that this effect is very weak even at extremes heliocentric angles \cite[see the results for the  K~{\sc i}~$\rm D_1$ line presented in][]{Uitenbroek1992}. Still, we study the PRD effects for both K~{\sc i} lines using the FALC model (in the previous work the VALC model \citep{Vernazza1981} was used instead). In this regard, we plot in Figure~\ref{prd1d} the results for the K~{\sc i}~$\rm D_1$ (left) and $\rm D_2$ (right) lines. For the latter case, we also included the neighbouring photospheric Fe~{\sc i} line (see label 4 in Table~\ref{lines}) as it is relatively close. We can see that, for all the selected heliocentric angles, the differences between the computed PRD and CRD profiles are negligible (see the bottom part of each plot). This is in agreement with the results of \cite{Uitenbroek1992} and indicates that PRD effects are very small for these lines, at least for the intensity spectrum.

\subsection{NLTE effects}

In previous works related to the alkali lines as in \cite{Bruls1992a} it has been demonstrated that the K~{\sc i}~$\rm D_1$ line, in particular its line core, is sensitive to NLTE effects. In other words, the LTE assumption is unable to reproduce the observed line depth. In this section, following the steps of the previous one, we check if the same occurs for the K~{\sc i}~$\rm D_2$ line and also revisit the studies of the K~{\sc i}~$\rm D_1$ line with the FALC model. In addition, as the results of the previous section revealed that PRD effects are negligible for the K~{\sc i} lines, we also compute the differences between NLTE and LTE for the polarization spectrum.

We show in Figure~\ref{nlte} the comparison between a NLTE and an LTE computation for the K~{\sc i} lines (the Fe~{\sc i} is always computed in LTE) assuming a heliocentric angle $\mu=1$. If we look at the first column, we can see that the line core in case of the NLTE computation (pink) is much deeper than for the LTE case (black). This happens for both potassium lines; the K~{\sc i}~$\rm D_2$ line is very deep with a line core intensity (in NLTE) of almost 0.1 of $I_c$. Moreover, if we examine the rest of the columns we also find differences between the NLTE and LTE polarimetric signals, the latter always with less amplitude. This is an interesting result that points out the necessity of a NLTE treatment for both the intensity and the polarization spectrum. Finally, we can see that these differences also affect to additional spectral features as the Stokes $V$ asymmetries. We plot in Figure~\ref{nlte_asymm} the Stokes $V$ LTE and NLTE profiles for both K~{\sc i} lines using the same FALC model but with a velocity gradient along the LOS ($\mu=1$) this time. The shape of the red (right) lobe, for both K~{\sc i} lines, is different between LTE and NLTE computations, what indicates that we should be careful if we plan to treat these spectral lines assuming the LTE approximation as it would provide unreliable values of the velocity and magnetic field vector.

\section{Results}

We have found that PRD effects are negligible for the K~{\sc i} lines. This lets us change to the CRD approximation, a necessary condition for computing the full Stokes spectra in RH. Therefore, from now on, all the studies we perform compute the full Stokes vector always under the CRD assumption and for a heliocentric angle $\mu=1$.

\subsection{Response functions}

We present in Figure \ref{2drf} the RF for the lines included in Table~\ref{lines} to changes in the temperature (top), line of sight (LOS) velocity (middle), and field strength (bottom). We followed the same method explained in \cite{QuinteroNoda2016} using the FALC model with a magnetic field of 1000~G constant with height, while the inclination and azimuth are equal to 45 degrees. Starting with the RF to changes in the temperature we can see that both K~{\sc i} lines are sensitive to that atmospheric parameter only at lower heights. In fact, these heights are lower than those sampled by other photospheric lines. If we focus on the RF for the LOS velocity we find a different behaviour, where the sensitivity of the K~{\sc i} lines reaches much higher layers, around $\log\tau\sim-4.5$. The same happens for the field strength (bottom panel). These heights are higher than in other photospheric lines and also than what we found for the photospheric lines of the 850~nm window in \cite{QuinteroNoda2017}. Thus, it seems that, although the sensitivity to the temperature is mainly in the lower photosphere, for the rest of the atmospheric parameters the line core is sensitive to changes in the upper photosphere. The reason for this is that, as the K~{\sc i} lines are scattering lines \citep{Uitenbroek1992}, their sensitivity to the temperature comes from lower layers, where their line source function is coupled to the local thermal conditions. In this regard, the RF to temperature depends, among other factors, on the derivative of the source function, that for these lines takes significant values at lower heights. However, their sensitivity to the velocity and the magnetic field comes from higher heights, closer to the location of less significant interaction with the atmosphere. This also happens to the Na~{\sc i} resonance lines \citep[for instance, ][]{Eibe2001} and for the O~{\sc i} infrared lines where, although their line intensity comes from the low photosphere, the scattering polarization signals are produced in the chromosphere \citep{delPinoAleman2015}.

\begin{figure}
\begin{center} 
 \includegraphics[trim=0 0 0 0,width=7.9cm]{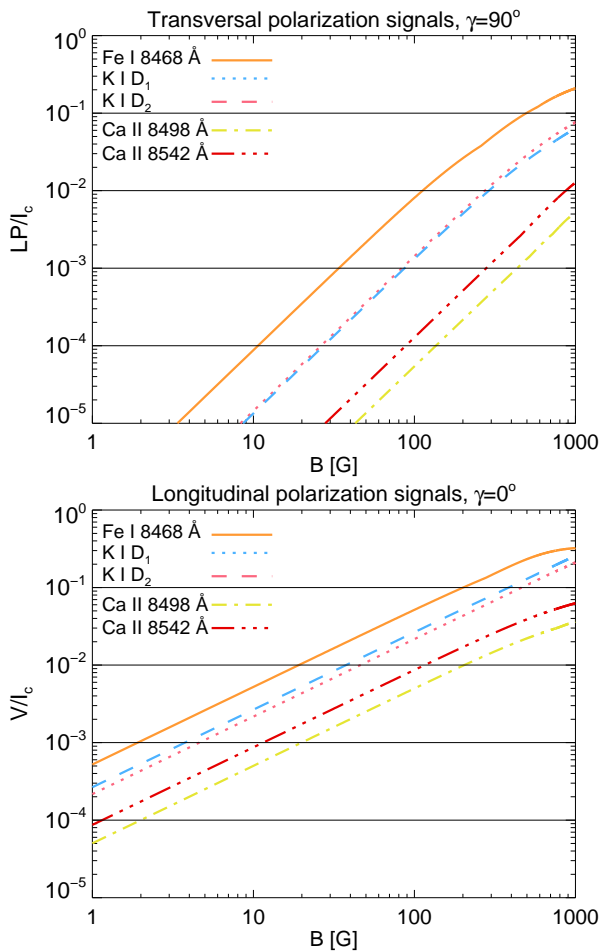}
 \vspace{-0.1cm}
 \caption{Maximum polarization signals for selected spectral lines from the 850~nm window and for the K~{\sc i} $\rm D_1$ (dashed blue) and $\rm D_2$ (dotted pink) lines. Upper panel shows the total linear polarization while the lower panel displays the maximum Stokes $V$ signals.}
 \label{1dpol}
 \end{center}
\end{figure}

\begin{figure*}
\begin{center} 
 \includegraphics[trim=18 0 35 0,width=15.2cm]{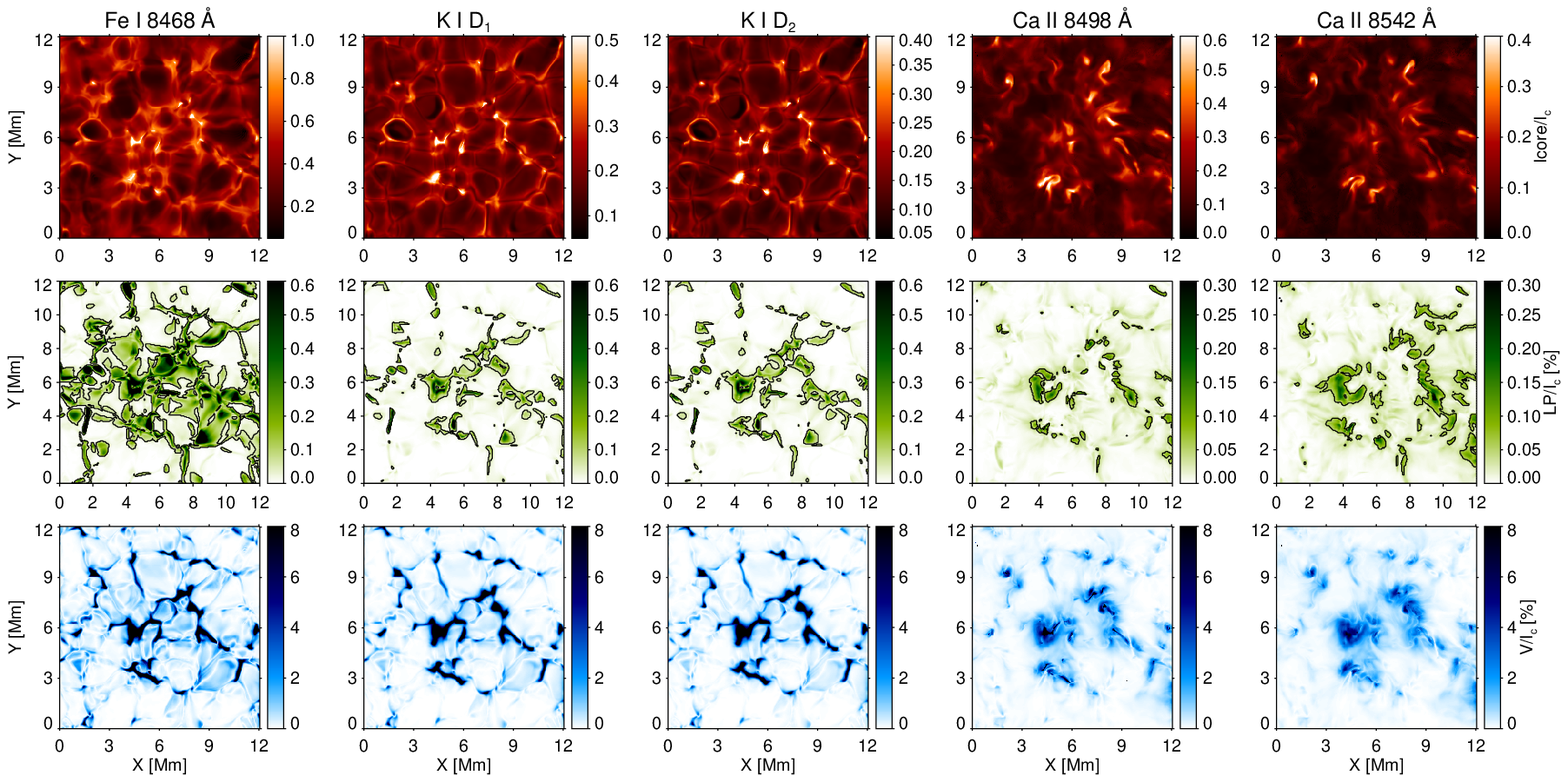}
 \vspace{+0.2cm}
 \caption{Spatial distribution of line core intensity (top), linear polarization (middle), and circular polarization signals (bottom) for selected lines. Starting from the left, Fe~{\sc i} 8468~\AA, K~{\sc i} $\rm D_1$, K~{\sc i} $\rm D_2$, Ca~{\sc ii} 8498~\AA \, and Ca~{\sc ii} 8542~\AA. Contours designate those locations with linear polarization signals larger than $5\times10^{-4}$ of $I_c$. The actual field of view corresponds with the one highlighted in Figure~\ref{fov}.}
 \label{polsignals}
 \end{center}
\end{figure*}

\subsection{Polarization signals}

In this section, we examine the maximum polarization signals for selected spectral lines and different atmospheric models. We first start with the FALC model used in the previous section and we include a magnetic field constant with height. Then we modify its field strength using 1~G steps from 1 to 1000~G. As we plan to estimate the transversal and longitudinal signals separately we perform the computation with an inclination value of 90~degrees first and, then, with 0~degrees. In both cases, the azimuth is fixed and equal to 45~degrees. In the first case (top panel of Figure~\ref{1dpol}), the K~{\sc i} linear polarization signals (computed as $LP=\sqrt{Q^2+U^2}$) are almost the same with values up to $10^{-3}$ of $I_c$ for a horizontal field of 100~G. The reason is that they remain under the weak field regime for the field strength values used in this study. In this regime, the linear polarization depends, among other factors, on the difference  $G_{0}^{(2)}- G_{1}^{(2)}$ \citep[see page 400 of][]{Landi2004}. If we compute this difference for the two potassium lines (see Table 4.3 and Equation 3.8 of the monograph) we obtain that $G_{0}^{(2)}- G_{1}^{(2)}=-1.33$ for both lines. Therefore, the linear polarization signals should be very similar if we are in the weak field regime. In addition, if we compare their signals with those produced by the rest of spectral lines, we find that they are in between those generated by the Fe~{\sc i} 8468~\AA \ and the Ca~{\sc ii} lines. 

Regarding the circular polarization (bottom panel of Figure~\ref{1dpol}), we have a similar behaviour with K~{\sc i} signals in between those produced by the photospheric and the Ca~{\sc ii} lines, reaching up to $10^{-2}$ of $I_c$ for a vertical field of 100~G. However, in this case, their amplitude is always slightly different, being the K~{\sc i} $\rm D_1$ line the one that produces larger signals. We believe that this is because the latter line has a larger Land\'{e} factor (see Table~\ref{lines}) that strongly affects the amplitude of the polarization signals in the weak field regime \citep[see Table 9.1 in the monograph of][]{Landi2004}.

We complement the previous study computing the maximum polarization signals using the Bifrost simulation. We focus on the region enclosed by the square in Figure~\ref{fov} and we perform column by column computations \citep[similar to what we did in][]{QuinteroNoda2016,QuinteroNoda2017} using the one dimensional  geometry package of RH. The results are presented in Figure~\ref{polsignals} where we also include the line core intensity (top row) to check if the results discovered in the RF to temperature are also found in a more complex atmosphere. Thus, if we examine the spatial distribution of the intensity in the core of the K~{\sc i} lines,  we detect the reverse granulation pattern, i.e. intergranular lanes are brighter than granules \citep[for instance,][]{Rutten2004}. There are also bright areas that correspond to the regions with a strong magnetic field. If we compare these results with the spatial distribution of signals for the Fe~{\sc i} 8468~\AA \ line, we can see that there are several similarities, but it seems that the pattern showed by the last one is more dynamic and the granulation structures are less defined. We believe that is probably because the photospheric line is sensitive to the temperature at higher heights, as we found in the previous section. For the case of the chromospheric lines we have that the spatial distribution of Ca~{\sc ii} infrared lines is completely different from both the photospheric and K~{\sc i} lines, indicating that the line core is sensitive to the temperature at higher atmospheric layers. 

Regarding the linear polarization, we have that both K~{\sc i} lines produce signals higher than the reference threshold of $5\times10^{-4}$ of $I_c$ (see contours) for several locations in the field of view. They are larger and occupy wider areas for the K~{\sc i} $\rm D_2$ line. If we compare these results with the photospheric iron line (first column) we can see that the pattern is relatively different because that one is more sensitive to the magnetic field \citep[see Table 1 of ][]{QuinteroNoda2017}. In the case of the Ca~{\sc ii} lines, they show again a different spatial distribution of signals, being located further from the magnetic field concentrations.

Concerning the circular polarization generated by the K~{\sc i} lines (bottom row), they are larger for the K~{\sc i} $\rm D_1$ line following the same trend found in the previous study (see Fig. \ref{1dpol}). If we compare them with the results for the iron line (first column), we find that the former occupy wider areas. This might indicate that these polarization signals come from higher heights in the photosphere, something that it is in agreement with the RF results presented in Figure~\ref{2drf}.

\subsection{Height of formation}

In this section we estimate the formation heights of the K~{\sc i} lines and compare with those of selected spectral lines. We add, for the first time in this work, an additional photospheric line, i.e. the Si~{\sc i} 7680~\AA \ absorption line (see label~6 in Table~\ref{lines} and Figure~\ref{2drf}). We aim to show how different could be the height of formation among photospheric lines, trying to reinforce the argument of measuring multiple lines with different heights of formation. We compute the height where the optical depth is unity for the line core wavelength of the spectral lines analysed. In order to understand how it depends on different atmospheric parameters, in particular on the magnetic field strength, we calculate the height of formation for a 2D slice of the snapshot computed using the {\sc bifrost} code (see the horizontal line in Figure~\ref{fov}). As the simulation includes LOS velocities that will shift the wavelength position of the line core we calculate this position for every pixel and every spectral line.

\begin{figure*}
\begin{center} 
 \includegraphics[trim=38 0 5 0,width=15.0cm]{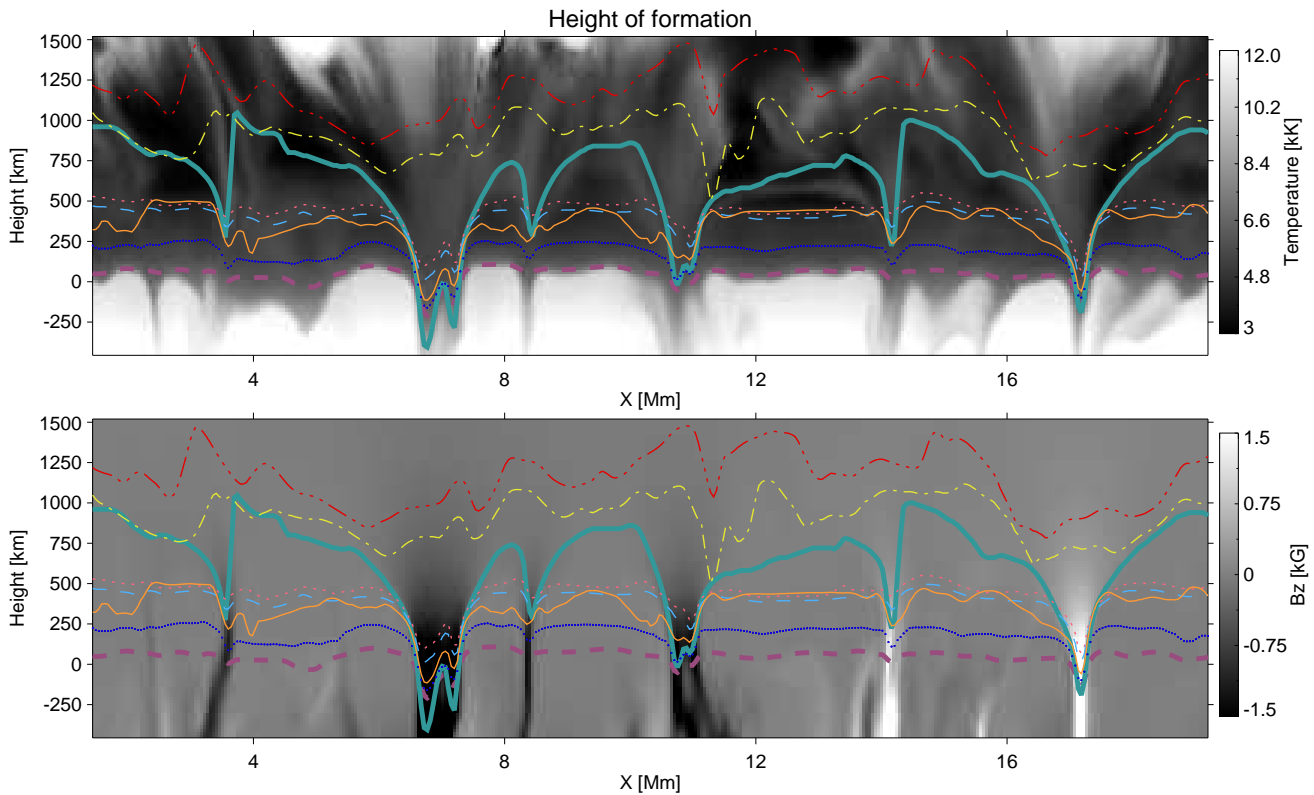}
 \vspace{+0.2cm}
 \caption{Height of formation of selected spectral lines. Each line (see also Figure~\ref{1dpol}) designates the height where the optical depth is unity for the line core wavelength of Si~{\sc i} 7680 \AA \ (blue dots), Fe~{\sc i} 8468~\AA \ (thin solid orange), K~{\sc i} $\rm D_1$ (thin dashed sky blue), K~{\sc i} $\rm D_2$ (dotted pink), Ca~{\sc ii} 8498 \AA \ (dash-dot yellow), and Ca~{\sc ii} 8542 \AA \ (dash-triple dot red). Background panels are the temperature (top) and longitudinal field (bottom) of the region indicated by the horizontal line in Figure \ref{fov}.  The thick solid green line designates the region where the plasma $\beta\sim1$ and the thick dashed purple line designates the height where optical depth is unity at 500~nm.}
 \label{height}
 \end{center}
\end{figure*}

Figure~\ref{height} shows the height of formation for different spectral lines, the height where the plasma $\beta$ (the ratio between the gas and the magnetic pressure) is unity (see solid thick line) as where the optical depth is unity at 500~nm (thick dashed purple line). Temperature (top) and longitudinal field (bottom) stratifications are included in the background as reference. The first thing we realize is that there are basically two distinct regions. The upper part, where the Ca~{\sc ii} lines form, always above or close to $\beta=1$, and the lower part where the photospheric Fe~{\sc i} and Si~{\sc i} lines and the K~{\sc i} lines form (usually below the height where $\beta=1$). In the case of the former, there are fluctuations in their height of formation but it is safe to assume that they form around 1000~km, with the Ca~{\sc ii} 8542 \AA \ line forming higher. For the latter lines, they usually form around 500~km (except for the weaker Si~{\sc i} line (blue dots) that forms below 250~km) when the magnetic field is weak and this height abruptly drops for larger concentrations of magnetic field. Interestingly, in those locations, the K~{\sc i} lines always form higher than the iron line (solid orange) with a difference varying between $100-250$~km, with the K~{\sc i}~$\rm D_2$ always formed the highest. This seems to reinforce the results of the previous section where the maximum circular polarization signals of the K~{\sc i}~$\rm D_2$ line occupied larger areas indicating that they are probably produced in a higher atmospheric layer where the magnetic field lines are expanded. Moreover, we can see that adding multiple lines, as we mentioned in \citet{QuinteroNoda2017}, we are covering different heights not only in the chromosphere but also in the photosphere (see the differences between the Si~{\sc i}, the Fe~{\sc i}, and potassium lines).

\section{Conclusions}

The main aim of this work was to characterize the K~{\sc i} $\rm D_1$ \& $\rm D_2$ lines in order to find if they could complement the capabilities of the spectral lines included in the 850~nm window studied in \cite{QuinteroNoda2017}. The results indicate that indeed they are sensitive to the LOS velocity and field strength at heights where neither the photospheric lines nor the Ca~{\sc ii} infrared lines of the mentioned window are. Therefore, if we observe both spectral windows, i.e. 770 and 850~nm, we would be seamlessly scanning different atmospheric layers from the bottom of the photosphere to the middle chromosphere (see, for instance, Figure~\ref{height}). However, we also found that their sensitivity to the temperature comes essentially from the photosphere. This is, unfortunately, something common for other resonance lines as the Na~{\sc i} lines. For instance, we can see in Figure 2 and 3 of \cite{Eibe2001} that the RF to perturbations in the LOS velocity of both Na~{\sc i} $\rm D_1$~\&~ $\rm D_2$ lines reach low chromospheric layers (around 800~km) while the RF to perturbations in the temperature does not go beyond  $300$~km above the continuum optical depth unity. Thus, it seems that it is difficult to find spectral lines that form, and are sensitive to the atmospheric parameters, at geometrical heights around 600~km (see Figure~\ref{height}). For this reason, we consider that the K~{\sc i} lines are good candidates for improving observations of the 850~nm window.

We plan to continue performing studies to understand how much we can gain from possible simultaneous observations of the 770~nm and 850~nm spectral windows. We plan to check the effect of PRD on the polarization Stokes profiles. In this work, we demonstrated \citep[although this was partially already done by ][]{Uitenbroek1992} that  both K~{\sc i} lines are insensitive to PRD effects for the intensity profiles but it would be interesting to check if the same happens for the polarization profiles. Some authors have done some effort in this direction \citep[for instance,][]{delPinoAleman2016,AlsinaBallester2017} for the Mg~{\sc ii} $h$ \& $k$, Sr~{\sc i}, and  Sr~{\sc ii} resonance lines and we would like to do a similar study for the K~{\sc i} lines. Moreover, we plan to investigate the amount of atomic polarization generated by these lines. Some works analysed the K~{\sc i} lines \citep[for example,][]{Belluzzi2007} for one-dimensional models that we would like to expand to a 3D computation assuming CRD following the steps of \cite{Stepan2016}. We believe that it will be interesting to examine how large are the atomic polarization signals at different heliocentric angles and how much the presence of a magnetic field alter them through the Hanle effect.

Finally, we want to close this work saying that, from both K~{\sc i} lines, the one that seems more interesting is the K~{\sc i} $\rm D_2$ because it forms slightly higher in the atmosphere. Something that also happens for the Na~{\sc i} $\rm D_2$ \citep[see][]{Eibe2001}. Thus, as this line cannot be accessed from the ground, we believe that observing it with a satellite or a balloon mission will open a new window of polarimetric observations, what will lead to a better understanding of the physical phenomena that takes place in the solar atmosphere.

\section*{Acknowledgements}
Special thanks to Valent\'{i}n Mart\'{i}nez Pillet, Sami Solanki and Tanaus\'{u} del Pino Alem\'{a}n  for their comments and suggestions. We would like to show our gratitude for the the anonymous referee that carefully revised the manuscript providing helpful comments and suggestions.  This work was supported by the funding for the international collaboration mission (SUNRISE-3) of ISAS/JAXA. The research leading to these results has received funding from the European Research Council under the European Union's Seventh Framework Programme (FP7/2007-2013)/ERC grant agreement no 291058. This work has also been supported by Spanish Ministry of Economy and Competitiveness through the project ESP2014-56169-C6-1-R. 

\bibliographystyle{mnras} 
\bibliography{potasium} 

\bsp	
\label{lastpage}
\end{document}